\documentclass[a4paper,10pt]{llncs}

\usepackage[colorlinks=true,urlcolor=blue,citecolor=black,linkcolor=red,final]{hyperref}
\usepackage[utf8]{inputenc}
\usepackage[english]{babel}
\usepackage{pdfcomment}
\usepackage[final]{pdfpages}
\usepackage[T1]{fontenc}
\usepackage{amssymb, xspace,multicol}
\usepackage{listings,amsmath,dsfont}
\usepackage{sectsty}
\usepackage{ifthen}
\usepackage{url}
\usepackage{caption}
\usepackage{framed}
\usepackage{courier}
\usepackage[]{algorithm2e}
\pdfoutput=1

\title{Synthesizing invariants by solving solvable loops}
\author{Steven de Oliveira$^1$, Saddek Bensalem$^2$, Virgile Prevosto$^1$}
\institute{1 : CEA, List~ 2 : Université Grenoble Alpes}

\lstdefinelanguage{algo}{
  morekeywords={def, if, else, non_det,while,do,done},
  morecomment=[s]{/*}{*/}
}

\newcounter{thm}
\newcounter{prop}

\newtheorem{Def}{Definition}

\newtheorem{Lemma}{Lemma}

\renewcommand{\k}{\ensuremath{\mathbb{K}}}
\newcommand{\kbar}{\ensuremath{\overline{\k}}}
\newcommand{\1}{\ensuremath{\mathds{1}}}

\newcommand{\ib}{\item[$\bullet$]}

\newcommand{\scalprod}[2]{\ensuremath{\left<#1,#2\right>}}

\newcommand{\pilatcite}{\cite{pilat_long}\xspace}

\newcommand{\toolalgo}{\textsc{PILA}\xspace}
\newcommand{\toolname}{\textsc{Pilat}\xspace}
\newcommand{\sagename}{\textsc{Sage}\xspace}

\newcommand{\aligator}{\textsc{Aligator}\xspace}
\newcommand{\fastind}{\textsc{Fastind}\xspace}

\newcommand{\pathtoprops}{}

\newcommand{\annexedProof}{1}
\newcommand{\longarticle}{1}

\newcommand{\labelprop}[1]{
  \ifthenelse{\inAnnex = 0}
  {\label{#1}}
  {} 

}

\newcommand{\propinput}[1]{
\ifthenelse{\annexedProof=0}
	   {\ifthenelse{\longarticle=0}
		       {\input{\pathtoprops#1}}
		       {\input{\pathtoprops#1} \input{\pathtoprops#1_proof}}}
	   {\ifthenelse{\inAnnex=0}
		       {\input{\pathtoprops#1}}
		       {\input{\pathtoprops#1} \input{\pathtoprops#1_proof}}}
      
}

  
\newcommand{\inAnnex}{0}

\begin{document}

\maketitle

\begin{abstract}
%

When proving invariance properties of a program, we face two problems.
The first problem is related to the necessity of proving tautologies of 
considered assertion language, whereas the second manifests in the need of 
finding sufficiently strong invariants. 
This paper focuses on the second problem and describes a new method for the 
automatic generation of loop invariants that handles polynomial and non 
deterministic assignments.
This technique is based on the eigenvector generation for a given linear 
transformation and on the polynomial optimization problem, which we implemented
in the open-source tool \toolname.

\end{abstract}

 \section{Introduction}
  Program verification relies on different mathematical foundations to
provide effective results and proofs of the absence of errors.
The problem is however undecidable for any Turing complete language,
partly because of loops.
This is one of the reasons why loop analysis is a highly studied topic in 
the field of verification.

Let us take for example linear filters, whose purpose is to apply a linear 
constraint to input signals.
This particular kind of programs is difficult to analyze because of the non 
determinism induced by the unknown input signal.
Here is an example of program inspired by linear filters: 
 \begin{lstlisting}
  x = non_det(-1,1);
  y = non_det(-1,1);
  while(x < 4) do
    N = non_det(-0.1,0.1);
    (x,y) = (0.68 * (x-y) + N, 0.68 * (x+y) + N);
  done
 \end{lstlisting}

%
We claim that loop invariants are a good solution in order to provide
general information about such a loop. 
In this particular case, the loop admits the invariant 
$x^2 + y^2 \leqslant 14.9$ bounding the maximal value of $|x|$ and $|y|$
to $3.9$ : this is an infinite loop.
More generally, if we can infer bounds for the value of the loop variables 
or for polynomial expressions of these variables we then are
able to perform precise analyses such as reachability analyses.

We aim at facing two major problems of numeric invariant generation,
namely the generation of polynomial relations between variables and
the search of inductive spaces in which variables of a program belong
to, in the context of simple (i.e. non-nested) loops composed of
polynomial and non deterministic assignments.
The relations we generate have the advantage to be completely independent from
the initial state of the loop, making them fully generic, as opposed to
full-program based techniques that start from a specific initial state.
This work is an extension of the algorithm \toolalgo introduced in~\pilatcite,
which generates polynomial equalities between variables manipulated by a 
simple deterministic loop.
We show in this paper that a refined version of this algorithm
can also produce inductive inequality invariants and tackle non-deterministic
assignments as well as deterministic ones.
Moreover, we add to this analysis an optimization algorithm enabling us to
minimize the inductive set described by invariants of non deterministic loops.

\paragraph{Contributions.}

The initial \toolalgo approach~\pilatcite generates inductive invariants as equality relations
(of the form $P(X) = 0$ with $P$ a polynomial). We extend this method
(Section~\ref{sec:overview}) to generate new kinds of inductive invariants
(of the form $P(X) \leqslant k$).
It is mostly based on linear algebra and is applicable to C programs
manipulating integers and floating point numbers ;
to simplify we describe the method on a simple imperative
language (Section~\ref{sec:setting}).
%
%
%
%
%
%
The two main results of this extension are the treatment of loops with deterministic 
(Section~\ref{sec:deter}) and non-deterministic assignments
(Section~\ref{sec:indeter}).
In the latter case, we reduce the problem of generating
invariants to the polynomial optimization problem.
An algorithm for solving this problem is given.
%
%
%
The proposed method in this paper is correct, fully 
implemented in \toolname and is currently part of the Frama-C 
suite~\cite{Kirchner2015} as an external open-source 
plug-in, available at~\cite{pilat_tool}.
We 
show its efficiency by applying it on 
several examples from related literature in section~\ref{sec:bench}.

\ifthenelse{\longarticle = 1}
{}
{Due to space constraints, proofs have been omitted. They are available 
in a separate report~\cite{pilat_nd_long}}.

 \section{Overview}\label{sec:overview}
  When synthesizing invariants, three ingredients are required : 
\begin{enumerate}
 \item what kind of invariants are computed ;
 \item what will be their most useful shape ;
 \item how strong they will be.
\end{enumerate}
In abstract interpretation for example, we first choose the type of
invariant that will be computed, i.e. the abstract domain, then a
symbolic execution of properties of this domain will shape the initial
state into an invariant that we will try to keep as strong as possible 
by applying appropriate widening and narrowing operators.
\paragraph{Deterministic case.}
Let us first recall how \toolalgo works on a simple example.
Consider the loop of figure~\ref{fig:simple_loop} for which we want to generate
all invariants (polynomials $P$ such that $P(x,y) = 0$) of degree $2$.
 \begin{figure}
  \begin{framed}
    \begin{lstlisting}
(x,y) = (non_det(-1,1),non_det(-1,1));
while(*) do
  (x,y) = (0.68 * (x-y), 0.68 * (x+y));
done
    \end{lstlisting}

  \end{framed}

  \caption{Simple affine loop}\label{fig:simple_loop}
  \end{figure}
Instead of starting with an initial state, which is not assumed to be known,
we generate relations that are preserved by each step of the loop.
Let $f$ be the loop transformation, (here 
$f(x,y) = (0.68 * (x-y), 0.68 * (x+y))$. 
A linear application $\varphi$ 
is a semi-invariant if, given any valuation of the variables,
it stays constant through one iteration of $f$.
In other words, it must respect the following property:

\begin{center}
  If $\varphi(X) = 0$ then $\varphi(f(X)) = 0$
\end{center}

In linear algebra, this is strictly equivalent to the following :

\begin{center}
  If $\varphi(X) = 0$ then $f^*(\varphi)(X) = 0$
\end{center}
where $f^*(\varphi) = \varphi \circ f$ is the dual application of $f$.
If there exists a scalar $\lambda$ such that 
$f^*(\varphi) = \lambda.\varphi$ (i.e. $\varphi$ an eigenvector of $f^*$ 
associated to the eigenvalue $\lambda$) the criterion becomes obviously true, 
thus $\varphi$ is a semi-invariant.
%
%

By enhancing the loop expressiveness with new variables representing
the value of the monomials of variables used in the loop, namely $x_2$ for 
$x^2$, $y_2$ for $y^2$ and $xy$ for $x*y$, we are also able to generate
polynomial relations.
Let us take for instance $x_2$.
As the new value of $x$ is  $0.68.(x-y)$, the new value of $x^2$ is
$0.68^2.(x^2-2.x.y + y^2)$. 
%
$x_2$ can then be expressed as a linear application of $x_2$, $xy$ and $y_2$.
%
%
%
More generally, any monomial of variables of the loop in 
figure~\ref{fig:simple_loop} evolves linearly along the execution of the 
enhanced loop.
A linear invariant generation technique for linear loops can 
generate polynomial invariants by using the newly introduced variables.

We have shown in~\pilatcite that the eigenvectors of $f^*$ are exactly the 
set of such invariants bound to the transformation $f$ but we only investigated 
precisely what happened for the eigenspace associated to $1$, which
returned affine invariants.
%
%
%
When the associated eigenvalue was not $1$ we provided some
methods in order to infer stronger invariants such as invariant 
simplification and removal of irrational invariants,
but the resulting relations were still too weak.
In the example of figure~\ref{fig:simple_loop}, the associated eigenvalue of 
the only invariant $x^2 + y^2$ is $0.9248$. 
We can conclude that $x^2 + y^2 = 0$ is inductive but if it does not 
respect the initial state, this is not an invariant.

The key idea of this paper is to consider not only equalities, but also 
inequalities.
If the left eigenvector $\varphi$ is associated to an eigenvalue $\lambda$ such that 
$ 0 < \lambda \leqslant 1$ then $\lambda.\varphi(X)$ will
necessarily be smaller than $\varphi(X)$. Thus
\begin{center}
  If $\varphi(X) \leqslant k$ then $f^*(\varphi)(X) \leqslant k $
\end{center}
is true, and $\varphi(X) \leqslant k$ is inductive.
In our example, the relation $x^2 + y^2 \leqslant k$ is inductive, and
contrarily to $x^2+y^2=0$ it can be made an invariant even if the initial values
of $x$ and $y$ are not $0$: we just have to choose $k=x_{init}^2+y_{init}^2$.

\paragraph{Non deterministic case.}

The same reasoning can be applied to treat non deterministic values in 
assignments.
By setting the non deterministic values to a random value, 
e.g. $0$, we are left to find inductive inequality relations, 
which can be easily performed as we just saw.
In the deterministic case, generated formulas are inductive 
because the set of possible values for $x$ and $y$ that respects the 
formula gets bigger by applying the loop transformation once.
%
%
Adding the non deterministic noise may lead to non inductive formulas.
A solution consists in finding upper and lower bounds for this noise and
check if the set obtained in deterministic case stays stable under this new
transformation.
If this is not the case, we must consider a weaker invariant.

  \section{Setting}\label{sec:setting}
  \paragraph{Mathematical background.}
  
  Given a field $\k$ with a total ordering $\leqslant$, $\k^n$ is the 
  vector space of dimension $n$.
  Elements of $\k^n$ are denoted $x = (x_1,...,x_n)^t$ a column vector.
  The variables vector of an application $f$ is denoted $X$.
  $\mathcal{M}_n(\k)$ is the set of 
  matrices of size $n*n$ and $\k[X]$ is the set of polynomials
  with coefficients in $\k$. 
  We note $\kbar$ the algebraic closure
  of $\k$, $\kbar = \{x. \exists P \in \k[X], P(x) = 0\}$.
  We will use $\scalprod{.}{.}$ the linear algebra standard notation, $\scalprod{u}{v} = u^t.v$,
  with $.$ the standard dot product. 
  The \emph{dual} of a linear application $f$ associated to the matrix $A$ 
  will be denoted $f^*$ and associated to the matrix $A^t$.
  The kernel of a matrix $A \in \mathcal{M}_n(\k)$, denoted $\ker(A)$, is the 
  vectorial space defined as $\ker(A) = \left\{ x \in \k^n, Ax = 0\right\}$.
  Every matrix of $\mathcal{M}_n(\k)$ admits a finite set of eigenvalues 
  $\lambda \in \kbar$ and their associated eigenspaces $E_\lambda$, defined as
  $E_\lambda = \ker(A - \lambda Id)$, where $Id$ is the identity matrix and 
  $E_\lambda \neq \{ 0\}$. 
  Similarly, every matrix $A$ admits \emph{left-eigenspaces}, i.e. eigenspaces of $A^t$.
  We denote $|.| : \kbar \rightarrow \k$ the modulus of an algebraic number and
  $\lVert.\rVert : \k^n \rightarrow \k$ the usual euclidean norm of a vector.
%
%
  The limit of a multivariate function $f : \k^n \rightarrow \k$ for 
  $\lVert X\rVert \rightarrow l$ is defined by the maximal value of $f(X)$
  with $\lVert X\rVert$ in the neighborhood of $l \in \k \cup \{+\infty\}$ 
  and be denoted $\lim\limits_{\lVert X\rVert\rightarrow l} f$.
%

%
\paragraph{Invariants.}
  A formula requires two canonical properties to be an invariant: it must be 
  true at the beginning of the loop (initialization); it must be preserved
  afterwards.
%
  Similarly to~\pilatcite, we define the inductive relation $\varphi$ by
  the following constraint: 
  \begin{Def}{Exact}\label{def:invar:ex}
   
 $\varphi \in \k^n$ is an \emph{exact inductive invariant} for an application
 $f$ iff
       
   \begin{equation}\label{eq:ex_invar}
  \forall X, |\scalprod{\varphi}{X}| = 0 \Rightarrow |\scalprod{\varphi}{f(X)}| = 0
  \end{equation}
  \end{Def}
    We add to this definition the concept of convergent and divergent inductive 
  relation :
  
  \begin{Def}{Convergence}\label{def:invar:conv}

  $\varphi\in\k^n$ is a \emph{convergent inductive invariant}
  for an application $f$ iff
  \begin{equation}\label{eq:conv_invar}
  \forall X,\forall k\in \k, |\scalprod{\varphi}{X}| \leqslant k \Rightarrow |\scalprod{\varphi}{f(X)}|\leqslant k
  \end{equation}
     
  \end{Def}
  
  \begin{Def}{Divergence}\label{def:invar:div}

  $\varphi\in\k^n$ is a \emph{divergent inductive invariant}
  for an application $f$ iff
  \begin{equation}\label{eq:div_invar}
  \forall X,\forall k\in \k
  |\scalprod{\varphi}{X}| \geqslant k \Rightarrow |\scalprod{\varphi}{f(X)}| \geqslant k
  \end{equation}

  \end{Def}

  A vector $\varphi$ satisfying the inductive relation is called a
  \emph{semi-invariant} in contrast with \emph{invariants}
  that also verifies the initialization criterion, denoted 
  $\scalprod{\varphi}{X_{init}}~\leqslant~k$ for convergent invariants 
  and $\scalprod{\varphi}{X_{init}}~\geqslant~k$ for divergent invariants 
  with $X_{init}$ the variables' initial values.  
  The exact semi-invariants set of a linear application $f$ is 
  the union of all eigenspaces of $f^*$ as proven in~\pilatcite.
  Also, we define the solvability of a mapping as introduced 
  in~\cite{rodriguez2007generating}.
  \begin{Def}\label{solv_map}
    Let $g\in(\k[X])^m$ be a polynomial mapping. $g$ is solvable if there exists 
    a partition of $X$ into sub-vectors of variables $x = w_1\uplus ...\uplus w_k
    $
    such that $\forall j.~1\leqslant j \leqslant k$ we have 
    \begin{equation*}
    g_{w_j}(x) = M_jw_j^t + P_j(w_1,...,w_{j-1},N)
    \end{equation*}
    
    with $P_j$ a polynomial and $N$ eventual non deterministic parameters.
  \end{Def}
\paragraph{Remark.}
We proved in~\cite{pilat_long} that deterministic solvable assignments are 
linearizable, i.e. they can be replaced by equivalent linear applications.
This allows us to consider linear mappings $X' = A.X$ with
$X$ a vector containing both variables and monomials of those 
variables to represent solvable assignments.
 \paragraph{Programming model.}

 We use a basic programming language whose syntax is given in 
 figure~\ref{fig:code_semantics}. 
$Var$ is the set of variables used 
by the program. 
 Variables take their value in a field $\mathbb{K}$. 
 A program state
 is then a partial mapping $Var \rightharpoonup \k$. 
 Any given program only uses a finite number $n$ of variables,
 thus program states can be represented as
 vectors $X = (x_1,...,x_n)^t$. 
 Finally, we assume that for all programs, there
 exists $x_{n+1} = \1$ a constant variable always equal to $1$.
 This allows to represent any affine assignment by a matrix.
      \begin{figure*}[thbp]
    \begin{framed}

    \begin{center}
     
    \begin{multicols}{2}
    \begin{tabular}{rcl}
      $i$ ::= && \xspace\xspace skip\\
      \xspace &&| $i;i$\\
      \xspace &&| $(x_1,..,x_n):=(exp_1,...,exp_n)$ \\
      \xspace &&| while $*$ do $i$ done\\
    \end{tabular}
    \columnbreak 

      \begin{tabular}{lcl}
      $exp$ ::= && \xspace\xspace $cst \in \mathbb{K}$\\
      \xspace &&| $x \in Var$\\
      \xspace &&| $exp + exp$ \\
      \xspace &&| $exp * exp$ \\
      \xspace &&| $non\_det(exp,exp)$
    \end{tabular}
    
    \end{multicols}
    \end{center}
  \end{framed}

    \caption{\label{fig:code_semantics} Code syntax}  

\end{figure*}
The expression $non\_det(exp_1,exp_2)$ returns a random value between the 
valuation of $exp_1$ and $exp_2$ when the program reaches this location.
Multiple variables assignments occur simultaneously within a single instruction.
We say an assignment is affine (resp. solvable) when its right values 
is an affine (resp. solvable) combination. 
Also, we say that an instruction is non-deterministic when it is
an assignment in which the right value contains the expression $non\_det$.

%
  
%
 \section{Convergent and divergent linear applications}\label{sec:intervals}
  \subsection{Deterministic assignments}\label{sec:deter}
   Being an inductive invariant requires for a formula $F$ to be true after an
 iteration of the loop under the hypothesis that $F$ holds before the iteration.
 The left eigenspace of a linear transformation 
 (i.e. the eigenspace of the transformation dual)
 is exactly its set of exact invariants as defined in
 definition~\ref{def:invar:ex}.
 \paragraph{Convergence.}
 By linear algebra
 \begin{equation}\label{eq:invar2}
 |\scalprod{\varphi}{X}|~ \leqslant k \Rightarrow |\scalprod{f^*(\varphi)}{X}| \leqslant  k
 \end{equation}
  is strictly equivalent to the definition~\ref{def:invar:conv} of convergent 
  semi-invariants.
%
 $|\scalprod{\varphi}{X}| \leqslant k $ represent what we call a \emph{domain described 
 by $\varphi$}, i.e. a polynomial relation.
 The previous constraint specify that the domain described by
 $\varphi$ is stable by $f$.

  The loop in figure~\ref{fig:simple_loop} admits the invariant 
  $x^2 + y^2 \leqslant 2$, a domain described by 
  $\varphi = (0,0,0,1,0,1)^t$ in the base $(\1,x,xy,x_2,y,y_2)$
  where $x_2$ represents $x^2$, $xy$ represents $x*y$ and $y_2$
  represents $y^2$.
%
%
  We can check with the \toolalgo algorithm that $\varphi$ is
  an exact semi-invariant 
  of the loop as it is a left eigenvector of the transformation performed by 
  the loop.
  As such, it generates a vectorial space of exact semi-invariants
  $I = \{k.(x^2 + y^2) = 0~|, k\in \k\}$, which is 
  a very poor result as $x^2 + y^2$ is constant only if it starts at $0$ (else, $k = 0$
  and we don't know anything about $x^2 + y^2$).
  We focus now on the eigenvalue associated to $\varphi$ on $f^*$, which is 
  $0.9248$.
  Thus, we can replace $|\scalprod{f^*(\varphi)}{X}|$ by $|\lambda|.|\scalprod{\varphi}{X}|$ in 
  (\ref{eq:invar2}), which returns : 
    \begin{equation*}
  |\scalprod{\varphi}{X}| \leqslant k \Rightarrow |\lambda|.|\scalprod{\varphi}{X}| \leqslant k
  \end{equation*}
  As $|\lambda| < 1$, the vector $\varphi$ satisfies the equation, thus 
  $\varphi$ is a convergent semi-invariant.
  Knowing the maximal initial value of $x^2 + y^2$ allows to determine the value of $k$, 
  which is $2$.
  More generally, we have :

 \input{\pathtopropsdomain} \begin{proof}
 
 If $|\lambda| \leqslant 1$, then $\varphi$ is a convergent semi-invariant
 (see introduction of section~\ref{sec:deter}).
 We will prove the following lemma: 
 
 \begin{Lemma}\label{lemma}
  $\left(\forall k, |\scalprod{\varphi}{X}| \leqslant k \Rightarrow |\scalprod{\varphi}{f(X)}|\leqslant k\right) \Rightarrow f^*(\varphi) = \lambda.\varphi$
 
 \end{Lemma}
  \begin{proof}
  With $k = 0$, we end up with the exact
  semi-invariant equation (\ref{eq:ex_invar}), whose solutions are eigenvectors of $f^*$.
  $\square$
  \end{proof}
 
 As the exact semi-invariants set of $f$ is the union
of the eigenspaces of $f^*$, 
 we can deduce that this set is a superset of all the relations
 satisfying (\ref{eq:conv_invar}). 
Moreover by the lemma~\ref{lemma}, we have 
\begin{equation*}
(|\scalprod{\varphi}{X}| \leqslant k \Rightarrow |\scalprod{\varphi}{f(X)}|\leqslant k) \Rightarrow (|<\varphi,X>| \leqslant k \Rightarrow |\lambda|.|<\varphi,X>| \leqslant k)
\end{equation*}

For $k = |<\varphi,X>|$ it is true if and only if $|\lambda| \leqslant 1$.
$\square$
%
%
  
%
%
\end{proof}

 \paragraph{Divergence.}
 
 The same reasoning applies for the generation of divergent invariants.
 For example, an eigenvalue $\lambda$ such that $|\lambda| > 1$ associated to a semi-invariant 
 $\varphi$ implies that $|\scalprod{\varphi}{X}| \geqslant k$ is an inductive invariant.
 Thus, we also have 
 
 \input{\pathtopropscodomain} \input{\pathtopropscodomain_proof}
 
 Note that this is only an implication this time.
 For example, the transformation $f(x,\1) = (x+\1,\1)$ admits 
 $x \geqslant x_{init}$ as a divergent invariant but the only
 left eigenvector of $f$ is $(0,1)$, which correspond to 
 the invariant "\emph{$\1$ is constant}".
Moreover, not all invariants of the form $P(X) \leqslant k$ are generated : 
the loop with the only assignment $x = x - 1$ admits the (non-convergent) 
invariant $x \leqslant x_{init}$. 
This invariant does not enter the scope of our setting as $|x| \leqslant x_{init}$
is false for $2x_{init} + 1$ iterations of $x = x-1$.
  
  \subsection{Non-deterministic assignments}\label{sec:indeter}

Some programs depend on inputs given all along their execution, for
example linear filters. 
More generally, an important part of program analysis consists in studying 
non-deterministic assignments. 
As an example let us consider the program in figure~\ref{fig:nd_loop},
a slightly modified version of the program in figure~\ref{fig:simple_loop}.
\begin{figure}
\begin{framed}
 
 \begin{lstlisting}
  while (*) do
    N = non_det(-0.1,0.1);
    (x,y) = (0.68 * (x-y) + N, 0.68(x+y) + N);
  done
 \end{lstlisting}

\end{framed}
 \caption{Non deterministic variant of the example~\ref{fig:simple_loop}}\label{fig:nd_loop}
\end{figure}

%
Our previous reasoning is not applicable now because,
due to the non-determinism of $N$,
the loop is no longer a linear mapping.

\paragraph{Idea.}
Intuitively, we will represent this loop by a matrix parametrized 
by $N$.
For that purpose we use the concept of abstract application introduced 
in~\cite{jeannet2014abstract}.

\begin{Def}\label{def:abs_mat}
 Let $I \subset \k$.
 An abstract linear application $f : \mathbb{I}^q \mapsto \mathcal{M}_n(\k)$ is an
 application associating a $q-\mathit{tuple} (N_1,...,N_{q}) \in I^q$ to a matrix.
 We will call the $\mathit{tuple}$ the \emph{parameter} of its image matrix by $f$, 
 and $f^*$ the dual application of $f$ (i.e. the application such that 
 $f^*(N) = (f(N))^T$).
 The expression of the parametrized matrix with respect to an abstract linear
 application
 will be called the \emph{abstract matrix}.
\end{Def}

 In our setting, the parameters are the non-deterministic values.
 For example, the previous loop can be represented by the abstract matrix 
 $M_N$ : 

 \[\left(\begin{array}{cccccc}
  1 & 0 & 0 & 0 & 0 & 0\\
  N & 0.68 & 0 & 0 & -0.68 & 0\\
  N^2 & 1.36N & 0 & 0.462 & 0 & -0.462\\
  N^2 & 1.36N & 0.925 & 0.462 & -1.36N & 0.462\\
  N & 0.68 & 0 & 0 & 0.68 & 0\\
  N^2 & 1.36N & 0.925 & 0.462 & 1.36N & 0.462
 \end{array}\right)\]

We have shown in section~\ref{sec:deter} that $M_0$ admits the invariant
$e_0=(0,0,0,1,0,1)$ associated to the eigenvalue $\lambda_0=0.9248$.
By decomposing $M_N$ as the sum of $M_0$ and $(M_N - M_0)$, we
also have $e_0.M_N = e_0.M_0 + e_0.(M_N - M_0) = \lambda_0.e_0 + \delta_0^N$,
where $\delta_0^N = e_0.(M_N - M_0) = (N^2,2.72N,0,0,0,0)$.
%
%
As the eigenvalue $\lambda_0$ is smaller than $1$, we are looking for 
relations $\varphi$ such that :

\begin{equation*}
\forall X, |\scalprod{\varphi}{X}| \leqslant k \Rightarrow  |\scalprod{M_N^T.\varphi}{X}| \leqslant k
\end{equation*}

We will call $e_0$ a \emph{candidate invariant} for $M_N$. For
$e_0$ to be an proper invariant for this transformation, the
following property must hold:

\begin{equation}\label{eq:non_det}
\forall X, |\scalprod{e_0}{X}| \leqslant k \Rightarrow |\lambda_0\scalprod{e_0}{X} + \scalprod{\delta_0^N}{X}| \leqslant k 
\end{equation}

Multiplying $|\scalprod{e_0}{X}|$ by $|\lambda_0|$ reduces its value.
We need to make sure that adding $\scalprod{\delta_0^N}{X}$
does not contradict the induction criterion by increasing 
the result over $k$.
\begin{figure}[htbp]  
\begin{framed}
 \includegraphics{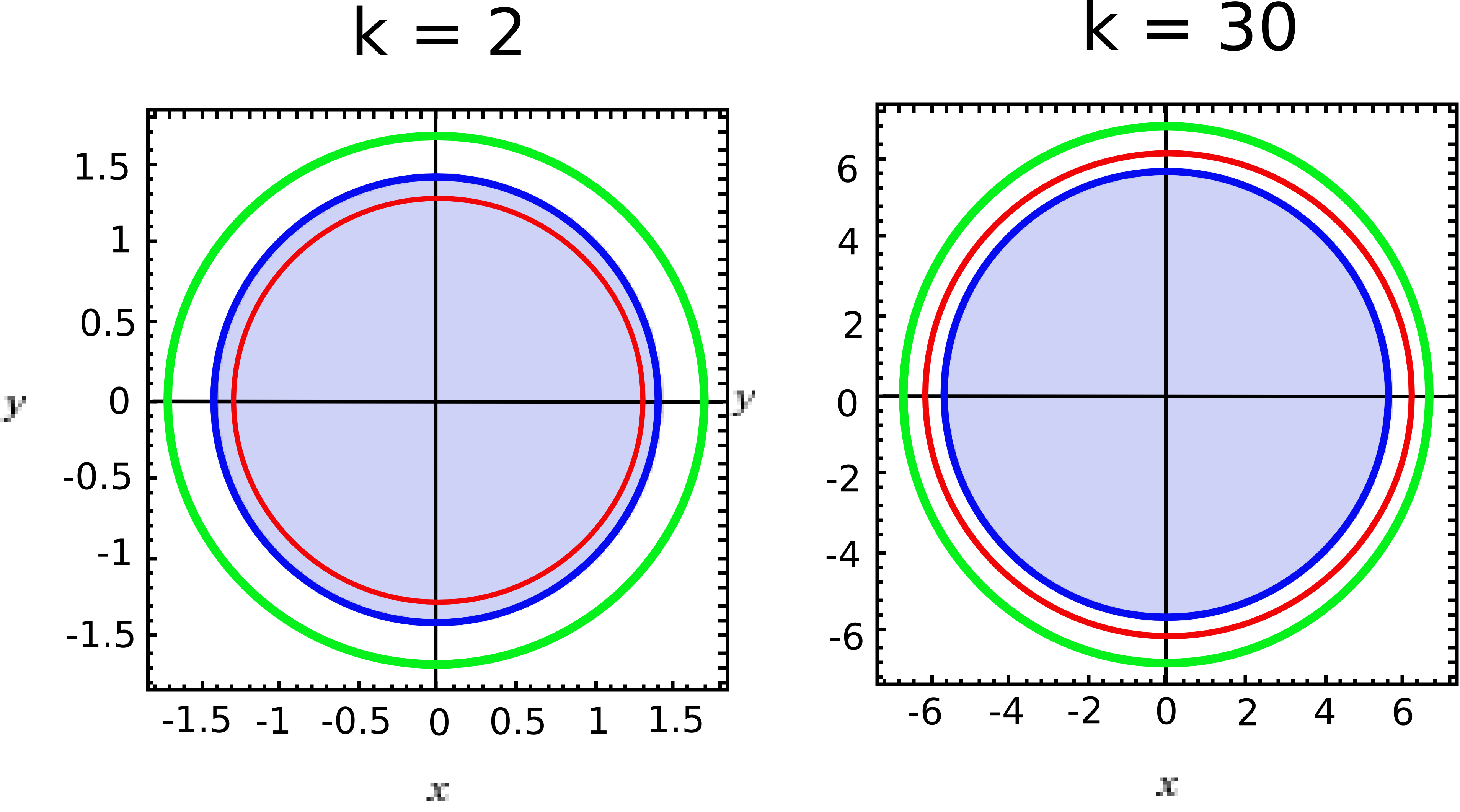}  
\end{framed}

 \caption{Representation of $x^2 + y^2 \leqslant k$ in blue, $|\lambda_0(x^2 + y^2)| \leqslant k$ in green and 
 $|\lambda_0(x^2 + y^2) + max(\scalprod{\delta_0^N}{X})| \leqslant k$ in red.}\label{fig:circle}
\end{figure}
We can see what happens on the figure~\ref{fig:circle}.
 When multiplied by a $\lambda < 1$, the value of $x^2 + y^2$ becomes smaller, 
 so the green circle $\lambda.(x^2 + y^2) \leqslant k$
 is bigger than the blue one (more values of $x$ and $y$ fits the equation). 

 \begin{itemize}
  \item[$\bullet$] In the first case $k$ is too small, adding $max(\scalprod{\delta_0^N}{X})$ 
  reduces the green circle too much. Thus, 
  the hypothesis that $(x,y)$ belongs to the blue
  circle does not imply it belongs to the red one:
  the candidate invariant is not inductive.
 
  \item[$\bullet$] In the second case, $max(\scalprod{\delta_0^N}{X})$ is too small to make the red
  back in the blue one: the candidate invariant is inductive.
 \end{itemize}
The variables of the program depend on $k$, as does $\scalprod{\delta_0^N}{X}$.
If it increases faster than $|\lambda_0\scalprod{e_0}{X}|$ when $k$
is increased, then no value of $k$ will make the candidate invariant inductive.
In particular,
if $\scalprod{e_0}{X}$ is a polynomial $P$ of degree $d$, we need 
to be able to give an upper bound of $\scalprod{\delta_0^N}{X}$ knowing that 
$|P(X)| < k$.
If the degree of $\scalprod{\delta_0^N}{X}$ is strictly 
smaller than $d$, then it will grow asymptotically slower than $|P(X)|$, thus
for a big enough $k$ the induction criterion is respected.
%


\propinput{non_det_induction_equiv}

%

%
%

%
%


%



We try to find a $k$ big enough for the set to be inductive. 
From the property~\ref{prop:nd_equiv_prop} we know that :
 \begin{equation*}
|\scalprod{\varphi}{X}| \leqslant k \Rightarrow -(1-\lambda_0).k\leqslant \scalprod{\delta_0^N}{X} \leqslant (1-\lambda_0).k
\end{equation*}

In our example, $\scalprod{\delta_0^N}{X} =  2.72*N*x + 2*N^2$. The 
polynomial $x$ is of degree $1$ while $<e_0,X> = x^2 + y^2$ is of degree 
$2$. 
%
%
We need to find a $k$ such that 

\begin{equation}\label{eq:objectives}
 -0.0752*k \leqslant 2.72*N*x + 2*N^2 \leqslant 0.0752*k
\end{equation}


%
\paragraph{Optimizing expressions.}

We need to maximize and minimize $2.72*N*x + 2*N^2$, knowing
the following three constraints:

\begin{itemize}
 \ib $x^2 + y^2 \leqslant k$
 \ib $N \leqslant 0.1$
 \ib $-0.1 \leqslant N$
\end{itemize}

Solving this problem is very close to solving a constrained polynomial
optimization (CPO) problem, 
a widely studied topic~\cite{bertsekas2014constrained}.
CPO techniques provide ways to find values minimizing and maximizing 
expressions constrained with inequalities between variables.
Our main issue is related to the parameter $k$ that must be known in order
to use CPO directly.
We will investigate in this article not how CPO works in detail, but how we can
reduce the problem of finding an optimal $k$ to the CPO problem, which enables 
us to use any CPO algorithm.

Assuming we have a function \emph{min} computing
the minimum, if it exists, of an expression under polynomial constraints, 
we propose an algorithm that refines the value of $k$ in figure~\ref{algo_k}.
\begin{figure}[htbp]
\begin{algorithm}[H]
 \KwData{~\\
    $\lambda$ : float\\
    $f$ : objective function\\ 
    $p$ : polynomial constraint\\ 
    non\_det\_c : non deterministic constraints\\ 
    $N$ : int}
 \KwResult{$k$ such that $\forall X, P(X)\leqslant k \Rightarrow f(X) \leqslant |(1-ev)|.k$}
 low\_k = 0\;
 up\_k = MAX\_INT\;
 k = MAX\_INT / 2\;
 i = 0\;
 \While{i<N}{
  i = i+1\;
  Q = (P(x) + k)\;
  min = min(f,[Q] + non\_det\_c)\;
  max = min(-1*f,[Q] + non\_det\_c)\;
  \eIf{min > (-1+ev) * k and max < (1-ev)*k}{
   up\_k = k\;
   }{
   low\_k = k\;
  }
  k = (low\_k + up\_k) / 2\;
 }
\end{algorithm}
 \caption{Dichotomy search of a $k$ satisfying the condition~(\ref{eq:objectives})}\label{algo_k}
\end{figure}
The idea is to find $k$ by dichotomy.
\begin{itemize}
  \ib If $k$ doesn't satisfy the constraints, we try a bigger one.
  \ib If we find a $k$ satisfying the two conditions over $k$, then
 it is a potential candidate. We can still try to refine it by searching
 for a $k$ slightly smaller.
\end{itemize}
We can improve this algorithm by guessing an upper value of $k$ instead of 
taking an arbitrary maximal value \emph{MAX\_INT}.
For our example, we started at $k = 50$ and found that $k = 14.9$ respects 
all the constraints. 

\begin{itemize}
 \ib $x^2 + y^2 \leqslant 14.9 \Rightarrow |x| \leqslant 3.9$ 
 \ib $|N| \leqslant 0.1$
\end{itemize}
Thus $|2.72*x*N + 2*N^2| \leqslant 1.0808$, and $k*0.0752 = 1.12$.

\paragraph{Remark.}
Note however that the existence of a $k$ satisfying~(\ref{eq:objectives}) is not guaranteed.
For example, the set 
$S = \{(x,y,N)| x^2 + y^2 \leqslant k \wedge -0.1 \leqslant N \leqslant 0.1\}$ is a compact 
set for any value of $k$, which means that $x$, $y$ and $N$ have maximum and 
minimum values.
This implies the existence of a lower and an upper bound for 
every expression composed with $x$, $y$ and $N$, but the value of those 
expressions may be always higher than $k$ such as for $x^2 + y^2 + 1$
bounded by $k+1$. 
\propinput{convergence}
By taking $M = (1-\lambda_0)$, 
this theorem gives us a sufficient condition to guarantee the convergence of 
the algorithm in figure~\ref{algo_k}.
As we are dealing with two polynomials $P$ and $Q$, then if $P$
(the candidate invariant) has a higher
degree than $Q$ (the objective function) in all its variables,
the limit of $\frac{Q(X)}{P(X)}$ will be
null which is enough to ensure the convergence of the method.
In our example, with $X=(x,y)$, $P(X) = x^2 + y^2$ and $Q(X,N) = 10.N(x^2 + y^2 + 1)$, 
with $|N| \leqslant 0.1$.
Because $\lim\limits_{\lVert X\rVert\rightarrow+\infty} |\frac{Q(X,N)}{P(X)} | = 10N$
is higher than $1$ for $N = 0.1$, the optimization procedure may not produce a result by 
theorem~\ref{convergence}.
In our case $Q(X) = 2,72.x.N + 2.N^2$ is a polynomial of degree $1$ in $x$ and 
$0$ in $y$, thus
$\lim\limits_{\lVert X\rVert\rightarrow+\infty} |\frac{Q(X,N)}{P(X)} | = 0$ 
and the optimization will converge.

\paragraph{Initial state.}
The knowledge of the initial state is not one of our hypotheses yet, but the 
previous theorem provides the necessary information we need to treat the case
where the initial state is strictly higher than the minimal $k$ we found.
The previous theorem tells us that there exists a $K$ such that for all $K'\geqslant K$,
$K'$ is a solution to the optimization problem. Our optimization algorithm
is searching for 
a value of $k$ for which the set is inductive, though, and
this solution may be only local : there may be a $k' > k$ which is not a solution of the
optimization procedure. 
If the value of $P(X_{init})$ is strictly higher than $k$, there are two 
possibilities : 
\begin{itemize}
 \ib it satisfies the objective (\ref{eq:objectives}) and this is a right
 solution (optimization is then not necessary as $k = P(X_{init})$ is correct);
 \ib it doesn't satisfy the objective and we have to find a $k$ higher than 
 $P(X_{init})$ satisfying it.
\end{itemize}

In both cases, we can enhance the optimization algorithm by first testing the 
objective~(\ref{eq:objectives}) with $k = P(X_{init})$.
If it does not respect the objective, then starting the dichotomy with 
$low\_k = P(X_{init})$ will return a solution (guaranteed by the 
theorem~\ref{convergence}) strictly higher than $P(X_{init})$.

  \subsection{Rounding error.}
  When dealing with real life programs, performing
floating point arithmetic generates rounding error.
%
%
As for an input signal abstracted by a non deterministic value, we can add to 
every computation that may lead to a rounding error a non deterministic 
value whose bounds are determined by the variables types and values.

\paragraph{Addition.}
Addition over two floating-point values lose some properties like 
associativity.
For example, $(2^{64} - 2^{64}) + 2^{-64} $ will be 
strictly equal to $2^{-64}$ but  $2^{64} + (- 2^{64} + 2^{-64})$ will be equal to $0$.
To deal with addition, we can consider the highest possible error 
between a real value and its floating point representation, a.k.a. the 
machine epsilon. 
It is completely dependent of the C type used :
for \emph{float} (single precision) it corresponds to $2^{-23}$ ;
for \emph{double} (double precision) it is $2^{-52}$.
More generally, let $x$ and $y$ be two reals, 
with $\tilde{x}$ and $\tilde{y}$ their 
respective C representation.
%
The IEEE standard model says that an operation on floating
point numbers must be equivalent to an operation on the 
reals, and then round the result to one of the nearest\footnote{depending on rounding mode, this may be the floating point value immediately below or above the result.} floating point number.
In this case, the relative error $|(\tilde{x} + \tilde{y}) - (x+y)| = (x + y)*\varepsilon$
where $\varepsilon$ is the highest machine epsilon between the machine epsilon 
of the type of $x$, $y$ and $(x+y)$.
The error is relative to the value of $x$ and $y$.
This is not a problem, as we authorize in our setting non deterministic calls with
expressions as argument. 

\paragraph{Multiplication.} 
A similar approximation happen during a multiplication of two floating point values.
%
%
The relative error $|(\tilde{x} * \tilde{y}) - (x*y)| = x*y*\varepsilon$
%
%
Thus for every multiplication, we can add a non deterministic value 
between $-x*y*\varepsilon$ and $x*y*\varepsilon$.

With these considerations, we are able to provide precise bounds for rounding error
for every operation performed in the loop.
\paragraph{Remark.}

Note that we also can deal with value casting.
For example, when a cast from a floating point value to a integer is performed,
the maximal error is bounded by $1$ which can be abstracted in our setting by 
a non deterministic assignment.

 \section{Related work}\label{sec:sota}
 There exist mainly two kinds of polynomial invariants: equality relations 
between variables, representing precise relations, and inequality relations, 
providing bounds over the different values of the variables.
After the results of Karr in~\cite{karr1976affine,muller2004note} on the 
complete search of affine equality relations between variables of an affine 
program, Müller-Olm and Seidl~\cite{muller2004precise} have proposed an 
inter-procedural method for computing polynomial equalities of bounded degree 
as invariants.
For linear programs, \emph{Farkas' lemma} can be used to encode the 
invariance condition~\cite{Colon2003} under non linear constraints.
Similarly, for polynomial programs, Gröbner bases have been shown to be 
an effective way to
compute the exact relation set of minimal polynomial loop invariants 
composed of \emph{solvable assignments} by computing the 
intersection of polynomial 
ideals~\cite{rodriguez2007generating,rodriguez2007automatic}.
Even if this algorithm is known to be EXP-TIME complete in the degree
of the invariant searched, high degree invariant is very rare for common loops
and the tool \aligator~\cite{kovacs2008aligator}, inspired from this technique
for \emph{P-solvable loops}\cite{kovacs2010complete,kovacs2008reasoning}, is 
very efficient for low degree loops. 
Finally, \cite{cachera2014inference} presents a technique that avoids the combination
problem by using abstract interpretation to generate \emph{abstract invariants}.
This technique is implemented in the tool \fastind.
The main issue is the completion loss: some invariants are missed and a 
maximal degree must be provided.
%

Synthesis of inequality invariants has become a growing 
field~\cite{mine2015algorithm,roux2012generic}, 
for example in linear filters analysis and automatic verification
in general as it provides good knowledge of the variables bounds when
computing floating point operations.
%
Abstract interpretation~\cite{cousot1977abstract} with 
widening operators 
allows good approximation of loops with the desired format.
A recent work~\cite{gonnord2014abstract} mixes abstract
interpretation and loop acceleration (i.e. the precise computation
of the transitive closure of a loop) to extend the framework and obtain
precise upper and lower bounds on variables in the polyhedron domain.
Very precise and computing non-trivial relations for complex loops and 
conditions, it has the drawback to be applicable to a very restricted type 
of transformations (linear transformation with eigenvalues $\lambda$ such that
$|\lambda| = 0$ or $1$). 
We see this technique as complementary to ours as it generates invariants we do
not find (such as $k \leqslant k_{init}$ for loop counters) and conversely.
In order to treat non-deterministic loops,
\cite{mine2015algorithm} refines as precisely as 
possible the set of reachable states for linear filters, harmonic 
oscillators and similar loops manipulating floating point numbers using a 
very specific abstract domain. 

The \toolalgo technique benefits from both of these domains as it is based
on the synthesis of precise relations over the variables of the 
program~\pilatcite and avoids using abstract interpretation so that invariants
have no predefined shape.
As some of those relations are \emph{convergent}(i.e. their valuation is 
reduced by every step of the loop) we can also deal with inequalities 
relations, and we provide a way to deal with non determinism with a
technique inspired by policy iteration~\cite{costan2005policy}.

 \section{Application and results}\label{sec:bench}
 The plug-in \toolname, written in OCaml as a Frama-C plug-in (compatible
with the latest stable release, Aluminium) and originally generating exact
relations for deterministic C loops, has been extended with convergent
invariant generation and non deterministic
loop treatment for simple C loops.
It implements our main algorithm of invariant generation in addition to the 
optimization algorithm of figure~\ref{algo_k}, and generates invariants
as ACSL~\cite{baudin2008acsl} annotations, making them readily understandable by
other Frama-C plugins.
The tool is available at~\cite{pilat_tool}.

\begin{figure}[htbp]
  \begin{framed}
  \begin{minipage}{0.57\textwidth}
    
    \begin{center}
      \vspace{-5ex}
    \end{center} 
    \begin{lstlisting}[language=C]
int simple_filter(int s0, int s1){
  float r;
  while (1) {
    r = 1.5*s0 - 0.7*s1 + [-0.1,0.1]
    s1 = s0;
    s0 = r;
  }
}
    \end{lstlisting}
  \end{minipage}
   \hfill
   \begin{minipage}{0.5\textwidth}
    \begin{center}
      \includegraphics[scale=0.6]{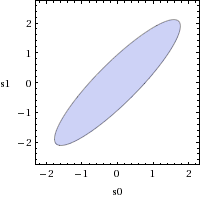}

    \end{center} 
   \end{minipage}

  \end{framed}
 \caption{\label{fig:pig_test} Generation of one of the smallest polynomial 
 invariant of degree 2 for a linear filter~\cite{mine2015algorithm,wolfram_picture}}
\end{figure}
Let us now detail the work performed by \toolname over the example of
figure~\ref{fig:pig_test} (taken from~\cite{mine2015algorithm}).
First, our tool generates the \emph{shape} of the invariant, i.e. the
polynomial $P$ such that $|P(X)| \leqslant k$ is inductive for a certain $k$ of 
the loop by setting the non deterministic choice to $0$. 
Here, the polynomial generated by \toolname is 
$P(s_0,s_1) = 1.42857*s_0^2 -2.14285*s_0*s_1 + s_1^2$ with the eigenvalue 
$\lambda = 0.7$.
Adding the noise to the matrix returns the noise polynomial 
$Q(s_0,s_1,N) = 2*N*s_1 - 2.142*N*s_0 - 1.428*N^2$ which has a lower degree 
than $P$ for a fixed $N$.
Thus, we have that
$\lim\limits_{\lVert (s0,s1)\rVert\rightarrow+\infty} \frac{Q(s0,s1,N)}{P(s0,s1)} = 0 < 1 - \lambda$.
The optimization procedure is now certain to converge, thus we minimize
and maximize $Q(X,N)$ with the hypothesis $P(s0,s1) \leqslant k$. 
%
%
%
%

%
%
By starting the procedure with $k = 50$ (which is usually a good heuristic) and 
performing $10$ iterations the optimization procedure returns $k =  0.87891$, thus 
$1.42857*s_0^2 -2.14285*s_0s_1 + s_1^2 \leqslant  0.87891$ is an inductive invariant.

Let us now consider that the initial state of the loop is $(s_0,s_1) = (2,1)$.
Then at the beginning of the loop, 
$1.42857*s_0^2 -2.14285*s_0s_1 + s_1^2 = 2.42858 > 0.87891$, which does not 
respect the invariant.
In this case the procedure starts by testing the optimization criterion 
with $k= 2.14285$. 
This choice of $k$ is correct.
In conclusion, we know that $1.42857*s_0^2 -2.14285*s_0s_1 + s_1^2 \leqslant 2.42858$
is an invariant of the loop.

More generally,
we evaluated our method over the benchmark used in~\cite{roux2013analyse}
for which we managed to find an invariant for every program containing no
conditions. 
Though this benchmark has been built to test the effectiveness of a specific 
abstract domain, we managed to find similar results with a more general 
technique.
Our results are given in table~\ref{table:results}.
\begin{table}[t!]

  \begin{center}
     \begin{tabular}{|c||c|c||c|c|c|}
	\hline
	 & & & & &\\
	Name & ~Var~ & ~Degree~ & \multicolumn{1}{c|}{\# invariants} & \multicolumn{1}{c|}{Candidate generation}& \multicolumn{1}{c|}{Optimization}\\
	 & & & & \multicolumn{1}{c|}{(in ms)} & \multicolumn{1}{c|}{(in s)}\\
	\hline
	\hline
	Simple linear filter & 2 & 2 & 1 & 1.5 & 1.3 \\
	\hline
	Simple linear filter & 2 & 4 & 5 & 21 & 18 \\
	\hline
	Example~\ref{fig:nd_loop} & 2 & 2 & 1 & 3 & 1.7\\
	\hline
	Linear filter& 4 & 2 & 1 & 1.9 & 1.5  \\
	\hline
	Lead-lag controller & 2 & 2 & 5 & 2.8 &11\\
	\hline
	Gaussian regulator & 2 & 2 & 1 & 7 & 2.5 \\
	\hline
	Controller & 4 & 1 & 2 & 1 & 5\\
	\hline
	Controller & 4 & 2 & 5 & 66 & 14\\
	\hline
	Low-pass filter & 5 & 2 & 4 & 60 & 7 \\
	\hline
	\hline
	Example~\ref{fig:simple_loop}&2 & 2  & 1 & 3 & --\\
	\hline
	Dampened oscillator & 4 & 2 & 1 & 7 & --  \\
	\hline
	Harmonic oscillator & 4 & 2 & 1 & 4 & --  \\
	\hline
      \end{tabular}
      
   \end{center}
 \caption{Performance results with our implementation \toolname. 
 Tests have been performed on a Dell Precision M4800 with 16GB RAM and 8 cores.
 The first part of the benchmark are non deterministic loops. 
 The second part represents deterministic loops (no optimization necesary).
 }\label{table:results}
 \end{table}
As ellipsoids are a suitable representation for those examples, we have choosen
$2$ as the input degree of almost all our examples.
The optimization script is based on~\sagename~\cite{stein2008sage}.
Note that the candidate generation is a lot faster than the optimization 
technique, mainly because of two reasons : 
\begin{itemize}
 \ib computing \emph{min} is time consumung for a big number of constraints;
 \ib it is imprecise and its current implementation is 
 incorrect (it outputs a lower approximation of the answer). We have to compute 
 verifications in order to find a correct answer.
\end{itemize}

 \section{Conclusion}\label{sec:conclusion}
  Invariant generation for non deterministic linear loop is known to be a 
difficult problem.
We provide to this purpose a surprisingly fast technique generating 
inductuve relations as it mostly relies on linear algebra algorithms widely 
used in many fields of computer science.
Also, the optimization procedure for the non determinism treatment returns
strong results.
These invariants will be used in the 
scope of Frama-C~\cite{Kirchner2015} as a help to static analyzers, weakest 
precondition calculators and model-checkers.
We are now facing three majors issues that we intend to address in the future:
the current optimization algorithm is assumed to have an exact \emph{min} 
function. 
However, such function is both time consuming and imprecise. 
In addition, 
conditions are treated non 
deterministically, which reduces the strength of our results and limits the
size of our benchmark to simple loops (linear filters with saturation 
are not included in our setting). Finally,
the search of invariants for nested loops is a complex problem on which
we are currently focusing.

\bibliography{main}
\bibliographystyle{abbrv}
  
  \ifthenelse{\longarticle = 1 \and \annexedProof = 1}
    {
    \newpage
    \section{Appendix}
	\renewcommand{\inAnnex}{1}
	\setcounter{thm}{0}
	\setcounter{prop}{0}
	
	%
%
%
  
  \subsection{Domain and codomains}
  
  The following justifies the two properties of section~\ref{sec:deter}
  
  \paragraph{Domain}
  \propinput{domain}
  
  \paragraph{Codomain} 
  \propinput{codomain}
  
  \subsection{Non-deterministic stability}
  
  The following justifies the stability condition property in section~\ref{sec:indeter}
  \propinput{non_det_induction_equiv}
  
  \subsection{Boundedness of polynomials}
  
  The following justifies the boundedness property of polynomials in section~\ref{sec:indeter}
  \propinput{convergence}

    }

\end{document}